\documentclass{IEEEtran}
\usepackage{amssymb}
\usepackage{amsfonts}
\usepackage{amsmath}
\usepackage{algorithm}
\usepackage{algorithmic}
\usepackage{tikz}
\usepackage{graphicx,subfigure,cite,multicol,multirow,diagbox,booktabs,array}
\usepackage{color}
\usetikzlibrary{shapes.geometric, arrows}
\ifCLASSINFOpdf
\else
\fi

\begin{document}
\title{Enhanced Secure Wireless Information and Power Transfer via Intelligent Reflecting Surface}
\author{Weiping Shi, Xiaobo Zhou, Linqiong Jia, Yongpeng Wu, \\Feng Shu,~\emph{Member},~\emph{IEEE}, and Jiangzhou Wang,~\emph{Fellow},~\emph{IEEE}
\thanks{Weiping Shi, Xiaobo Zhou, Linqiong Jia and Feng Shu are with School of Electronic and Optical Engineering, Nanjing University of Science and Technology, Nanjing, 210094, China. (e-mail: shufeng@njust.edu.cn).}
\thanks{Yongpeng Wu is with the Shanghai Key Laboratory of Navigation and Location Based Services, Shanghai Jiao Tong University, Minhang 200240, China. (e-mail: yongpeng.wu2016@gmail.com).}
\thanks{Jiangzhou Wang is with the School of Engineering and Digital Arts, University of Kent, Canterbury CT2 7NT, U.K. (e-mail: j.z.wang@kent.ac.uk).}}
\maketitle
\begin{abstract}
In this paper, secure wireless information and power transfer with intelligent reflecting surface (IRS) is proposed for a multiple-input single-output (MISO) system. Under the secrecy rate (SR) and the reflecting phase shifts of IRS constraints, the secure transmit beamforming at access point (AP) and phase shifts at IRS are jointly optimized to maximize the harvested power of energy harvesting receiver (EHR). Due to the non-convexity of optimization problem and coupled optimization variables, firstly, we convert the optimization problem into a semidefinite relaxation (SDR) problem and a sub-optimal solution is achieved. To reduce the high-complexity of the proposed SDR method, a low-complexity successive convex approximation (SCA) technique is proposed. Simulation results show the power harvested by the proposed SDR and SCA methods approximately double that of the existing method without IRS given the same SR. In particular, the proposed SCA achieves almost the same performance as the proposed SDR but with a much lower complexity.
\end{abstract}


\IEEEpeerreviewmaketitle
\section{Introduction}
Sustainable green, cost-effective and secure techniques are basic requirements for beyond fifth-generation (5G) and sixth-generation (6G) communication system\cite{wu_TWC}. As a high energy-efficient tool, intelligent reflecting surface (IRS), which may adjust the phase shifts automatically via a large number of low-cost, passive and reflecting units\cite{Basar_IRS}, has recently attracted wide research attention from both industry world and academia.

There have been some innovative studies on wireless communication system with IRS by jointly optimizing the beamforming and phase shifts at the IRS \cite{wu_one,Energy_Efficiency,GUO_weight_sum_rate}. In a single-user multiple-input single-output (MISO) scenario \cite{wu_one}, in order to maximize the total received signal power, semidefinite relaxation (SDR) and Gaussisn random algorithms were
proposed to obtain a sub-optimal solution. Similarly, in the case of multi-user MISO \cite{Energy_Efficiency} applied gradient descent and sequential fractional programming to maximize the energy efficiency. Particularly, continuous and discrete  phase shifts were exploited in \cite{GUO_weight_sum_rate}, and the authors proposed Lagrangian dual transform to decouple the coupled optimization variables.

On the other hand, due to the fact that physical layer security technique could improve the wireless communication security significantly\cite{chenXM_secury,wangMH_secury,zhangN_secury,zhou_UAV}, secure wireless communication assisted by IRS has been investigated in\cite{cui_security,SHEN_Secrecy_IRS} to enhance the achievable secrecy rate (SR) of the legitimate user. An sub-optimal transmit beamforming vector and phase shifts vector of IRS have been designed for MISO secrecy channels aided by IRS with single-antenna eavesdroppers, in which majorization-minimization (MM) technique was proposed to optimize phase shifts and simplify calculation.

Moreover, simultaneous wireless information and power transfer (SWIPT) could enhance the energy efficiency and solve energy-limited issues of wireless networks to some extent. And secure wireless information and power transfer beamforming were investigated in\cite{XU_SWIPT_SEC}, where the power receivers were viewed as potential eavesdroppers, then the maximum secrecy rate and weighted sum-power were optimized. However, due to the severe path loss, wireless power transfer was only suitable for short distance transmission, hence the range of energy harvesting receivers (EHR) is limited. To enhance the harvested power of EHR, the SWIPT wireless network assisted by IRS has been researched. In \cite{WU_SWIPT_IRS}, the authors made an investigation of the maximization problem of weighted sum power for IRS aided SWIPT, and proved that it was not necessary to send a dedicated energy beamforming to EHR, where SDR technique was applied to solve the problem. Moreover, the study in \cite{PAN_IRS_MIMO} extended the case to MIMO broadcasting channels with the purpose of maximizing the weighted sum-rate.

To the best of our knowledges, secure IRS-aided SWIPT  has not still been reported. Due to the broadcast characteristics of wireless communication, physical layer security should be considered in SWIPT. In this paper, we maximize the harvested power of EHR subject to the secrecy transmission and uni-modular phase shifts of the IRS constraints. Our main contributions are as follows:
\begin{enumerate}
  \item A secure SWIPT system model assisted by IRS is proposed, in which a multi-antenna access point (AP) is to serve an EHR and an information receiver with the help of an IRS in the presence of an eavesdropper. By jointly optimizing the secure transmit beamforming and the phase shifts at the IRS, the harvested power maximization problem (HPMP) is established for an IRS-aided secure MISO-SWIPT system.
  \item Due to non-convex secure rate constraint, the uni-modular phase shifts constraint, and the coupled  optimization variables, the HPMP is non-convex and infeasible. To address this issue, the problem is solved via alternating optimization (AO) algorithms. Firstly, the problem is converted into a linear optimization problem by applying the trace function. Secondly, SDR-based AO and Gaussian randomization methods are adopted to get a sub-optimal solution.
  \item However, the above proposed SDR method has a high complexity. To reduce its computational complexity, an SCA-based AO algorithm is proposed. By utilizing the first-order Taylor approximation and inequality transformation, a sub-optimal solution is also obtained. Moreover, the phase shifts of the IRS are computed in semi-closed-forms per iteration. Simulation results show that our proposed SDR and SCA algorithms can harvest higher power than conventional schemes without IRS.
\end{enumerate}

Notations: Lowercase letters represent scalars. Boldface uppercase and lowercase letters stand for matrices and vectors, respectively. $|\cdot|$ and $\|\cdot\|$ denote the  modulus of a scalar and Euclidean norm of a vector, respectively. Signs $\Re(\cdot)$ and $\arg(\cdot)$ represent the real part and the phase of a complex number. $(\cdot)^H$, $(\cdot)^*$, $\mathbb{E}$ and $\mathrm{tr}(\cdot)$ represent conjugate transpose, conjugate, expectation and trace of a matrix, respectively.

\section{System Model}
\begin{figure}[htb]
  \centering
  \includegraphics[width=0.48\textwidth]{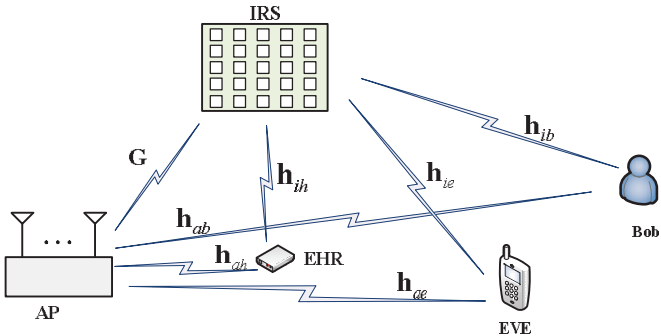}
  \caption{An IRS-aided secure SWIPT wireless network.}\label{Sys_Mod}
  \label{sys}
\end{figure}

Fig.~\ref{Sys_Mod} sketches a downlink MISO system with an IRS for SWIPT. In Fig.~\ref{Sys_Mod},  there are an AP with $M$ transmit antennas, an IRS with $N$ reflecting units, an information receiver  denoted as Bob,  and an EHR in the presence of an eavesdropper (EVE). All receivers are equipped with single antenna. The transmit signal from AP can be expressed as
\begin{equation}\label{Tx signal s}
\mathbf{x}=\mathbf{w}s,
\end{equation}
where $\mathbf{w}\in\mathbb{C}^{M\times 1}$  denotes the transmit beamforming vector, which forces the confidential message (CM) to the desired direction, and $\mathbf{s}\sim\mathcal{C}\mathcal{N}(0,1)$ is the CM for the Bob. Suppose that $P_s$ is the total transmission power constraint. Thus, we have $\mathbb{E}\left\{\mathbf{x}^H\mathbf{x}\right\}=\|\mathbf{w}\|^2\leq P_s$

In this paper, we assume a quasi-static fading environment. The baseband equivalent channel responses from
AP to IRS, from AP to Bob, from AP to EHR, from AP to EVE, from IRS to Bob, from IRS to the EHR, from IRS to EVE are denoted by $\mathbf{G}\in\mathbb{C}^{N\times M}$, $\mathbf{h}_{ab}^H\in\mathbb{C}^{1\times M}$, $\mathbf{h}_{ah}^H\in\mathbb{C}^{1\times M}$, $\mathbf{h}_{ae}^H\in\mathbb{C}^{1\times M}$, $\mathbf{h}_{ib}^H\in\mathbb{C}^{1\times N}$, $\mathbf{h}_{ih}^H\in\mathbb{C}^{1\times N}$, $\mathbf{h}_{ie}^H\in\mathbb{C}^{1\times N}$, respectively. The diagonal reflection-coefficient matrix of the IRS is denoted as $\mathbf{\Theta}=diag(\mathbf{\beta}_n\mathbf{e}^{j\mathbf{\theta}_{1}},\cdots,\mathbf{\beta}_n\mathbf{e}^{j\mathbf{\theta}_{N}})$, where $\theta_n\in [0,2\pi)$ and $\beta_n\in (0,1]$\cite{wu_one}. $\theta_n$ and $\beta_n$ are the phase shift and amplitude
reflection-coefficient of the $n$th unit, respectively. In this paper, $\beta=1$. The receive signal at Bob can be written as
\begin{align}\label{Rx_signal yb}
y_b(\mathbf{w},\mathbf{\Theta})
&=(\mathbf{h}_{ib}^H\mathbf{\Theta}\mathbf{G}+\mathbf{h}_{ab}^H)\mathbf{w}s+n_b,
\end{align}
where $n_b\sim\mathcal{C}\mathcal{N}(0,\sigma_b^2)$ is the complex additive white Gaussian noise (AWGN). Similarly, the received signals at EVE and at EHR are
\begin{align}
y_{e}(\mathbf{w},\mathbf{\Theta})
&=(\mathbf{h}_{ie}^H\mathbf{\Theta}\mathbf{G}+\mathbf{h}_{ae}^H)\mathbf{w}s+n_e,\label{Rx_signal ye}\\
\label{Rx_signal yr}
y_{r}(\mathbf{w},\mathbf{\Theta})
&=(\mathbf{h}_{ih}^H\mathbf{\Theta}\mathbf{G}+\mathbf{h}_{ah}^H)\mathbf{w}s+n_h,
\end{align}
respectively, where $n_e$ and $n_h$ are the complex AWGN variables,  which follow the distribution $n_e\sim\mathcal{C}\mathcal{N}(0,\sigma_e^2)$ and $n_h\sim\mathcal{C}\mathcal{N}(0,\sigma_h^2)$. Moreover, we assume that $\sigma_b^2=\sigma_e^2=\sigma_h^2=\sigma^2$. According to (\ref{Rx_signal yb}) and (\ref{Rx_signal ye}), the achievable transmission rate at Bob and EVE can be expressed as\cite{cui_security}
\begin{align}\label{Rb}
R_b(\mathbf{w},\mathbf{\Theta})
&=\log_2\left(1+\frac{|(\mathbf{h}_{ib}^H\mathbf{\Theta}\mathbf{G}+\mathbf{h}_{ab}^H)\mathbf{w}|^2}
{\sigma^2}\right)
\end{align}
and
\begin{align}\label{Re}
R_{e}(\mathbf{w},\mathbf{\Theta})
&= \log_2\left(1+\frac{|(\mathbf{h}_{ie}^H\mathbf{\Theta}\mathbf{G}+\mathbf{h}_{ae}^H)\mathbf{w}|^2}
{\sigma^2}\right),
\end{align}
respectively. The corresponding achievable secrecy rate (SR) is defined by \cite{wuXM_Secury}
\begin{align}\label{Rs}
R_s(\mathbf{w},\mathbf{\Theta})
&=\max\left\{0,R_b(\mathbf{w},\mathbf{\Theta})- R_{e}(\mathbf{w},\mathbf{\Theta})\right\}.
\end{align}

In addition, because of the broadcast nature of wireless channels, the energy is carried by information beam. The power harvested at EHR is \cite{XU_SWIPT_SEC}
\begin{equation}\label{E_h}
E_r(\mathbf{w},\mathbf{\Theta})=\zeta(|(\mathbf{h}_{ih}^H\mathbf{\Theta}\mathbf{G}+\mathbf{h}_{ah}^H)\mathbf{w}|^2),
\end{equation}
where $\zeta$ denotes the efficiency of power harvesting.

\section{Problem Formulation and Proposed Solution}
In this section, we maximize the harvested power at EHR by jointly optimizing the secure transmit beamforming vector and phase shifts at IRS to ensure that the achieved SR is greater than a predefined threshold. Moreover, the channel state informations of all the receivers' channels is assumed to be available at AP and IRS. Then the optimization problem can be mathematically cast as
\begin{subequations}\label{P1}
\begin{align}
\mathrm{(P1):}&\max_{\mathbf{w},\mathbf{\Theta}}~~~ E_r(\mathbf{w},\mathbf{\Theta})\\
&~~\text{s. t.}~~R_s(\mathbf{w},\mathbf{\Theta})\geq r_0\\
&~~~~~~~~\|\mathbf{w}\|^2 \leq P_s \\
&~~~~~~~~0 \leq \theta_n \leq 2\pi, \forall n=1\cdots N,
\end{align}
\end{subequations}
where $r_0>0$ denotes the minimum SR, $P_s>0$ refers to the prescribed power budget at AP and $\theta_n$ is the phase shifts of the elements at the IRS. It can be observed that problem $\mathrm{(P1)}$ is non-convex because the objective function and the constraints are non-convex as well as optimization variables $\mathbf{w}$ and $\mathbf{\Theta}$ are coupled. It is particularly noted that the objective function is convex  with respect to $\mathbf{w}$ and $\mathbf{\Theta}$, respectively. Thus we address the corresponding optimization problem by applying the alternating and iterative manner in the following.

Defining $\mathbf{u}=[\mathbf{e}^{j\theta_{1}},\cdots,\mathbf{e}^{j\theta_{N}}]^H$,
 $\mathbf{v}=[\mathbf{u};1]$, $\mathbf{H}_r=[\mathrm{diag}\{\mathbf{h}_{ih}^H\}\mathbf{G}; \mathbf{h}_{ah}^H]$, $\mathbf{H}_b=[\mathrm{diag}\{\mathbf{h}_{ib}^H\}\mathbf{G}; \mathbf{h}_{ab}^H]$, and $\mathbf{H}_e=[\mathrm{diag}\{\mathbf{h}_{ie}^H\}\mathbf{G}; \mathbf{h}_{ae}^H]$, problem $\mathrm{(P1)}$ is equivalent  to
\begin{subequations}\label{P2}
\begin{align}\label{P2objec}
\mathrm{(P2):}&\max_{\mathbf{w},\mathbf{v}}~~~ |\mathbf{v}^H\mathbf{H}_r\mathbf{w}|^2\\\label{P2_Rs}
&\text{s. t.}~~|\mathbf{v}^H\mathbf{H}_b\mathbf{w}|^2+\sigma^2\geq 2^{r_0}(|\mathbf{v}^H\mathbf{H}_e\mathbf{w}|^2+\sigma^2)\\
&~~~~~~\|\mathbf{w}\|^2 \leq P_s \\
&~~~~~~|\mathbf{v}_n|=1, \forall n=1\cdots N~~~\mathbf{v}_{N+1}=1.\label{orginal_v}
\end{align}
\end{subequations}
\subsection{Proposed SDR-based alternating optimization method}\label{subA}
In this subsection, we present a near optimal solution to problem $\mathrm{(P2)}$. For brevity, we rewrite (\ref{P2objec}) as $f(\mathbf{W},\mathbf{V}) \triangleq \mathrm{tr}(\mathbf{H}_r^H\mathbf{V}\mathbf{H}_r\mathbf{W})$, and (\ref{P2_Rs}) as
\begin{align}\label{Rs_constraint}
\mathrm{tr}(\mathbf{H}_b^H\mathbf{V}\mathbf{H}_b\mathbf{W})+\sigma^2
\geq 2^{r_0}(\mathrm{tr}(\mathbf{H}_e^H\mathbf{V}\mathbf{H}_e\mathbf{W})+\sigma^2),
\end{align}
where $\mathbf{V}=\mathbf{v}\mathbf{v}^H$  and $\mathbf{W}=\mathbf{w}\mathbf{w}^H$.
Therefore, after dropping the rank-one constraints $\text{rank}(\mathbf{W})=1$ and $\text{rank}(\mathbf{V})=1$, the SDR of problem $\mathrm{(P2)}$ is
\begin{subequations}\label{P3}
\begin{align}
\mathrm{(P3):}&\max_{\mathbf{W},\mathbf{V}}~~~ f(\mathbf{W},\mathbf{V})\\
&~~\text{s. t.}~~ \mathrm{tr}(\mathbf{W})\leq P_s\label{tr_W}\\
&~~~~~~~~\mathrm{tr}(\mathbf{E}_n\mathbf{V})=1, \forall n=1\cdots N+1\label{tr_V}\\
&~~~~~~~~\mathbf{W}\succeq 0,\mathbf{V}\succeq 0,(\ref{Rs_constraint}),
\end{align}
\end{subequations}
where $\mathbf{E}_n\in \mathbf{R}^{(N+1)\times (N+1)}$, and $\mathbf{E}_n$ means that the value of the element on position $(n,n)$ is 1 and 0 otherwise. Due to the fact that $\mathbf{W}$ and $\mathbf{V}$ are coupled and problem $\mathrm{(P3)}$ is non-convex, it is difficult to solve this  kind of non-convex problems directly. However, problem $\mathrm{(P3)}$  could be decomposed  into two subproblems and solved by applying AO algorithm. By alternately fixing $\theta$ and $\mathbf{w}$, $\mathrm{(P3)}$ is reduced to two standard semidefinite programs (SDP), which can be solved by CVX directly. Making  use of the AO algorithm, we may obtain the solution to problem $\mathrm{(P3)}$. However, because problem $\mathrm{(P3)}$ ignores rank-one constraints $rank(\mathbf{W})=1$ and $rank(\mathbf{V})=1$, to recover them, we apply standard Gaussian randomization method and obtain a high-quality sub-optimal solution of problem $\mathrm{(P2)}$, the detail is similar to that of \cite{wu_one}. In addition, the objective value of problem $\mathrm{(P3)}$ is non-decreasing after each iteration, therefore, the SDR-based AO algorithm is guaranteed to converge to a locally optimal solution.
\subsection{Proposed low-complexity SCA method}
In subsection \ref{subA}, we have proposed the SDR-based AO Algorithm to obtain the information beamforming matrix $\mathbf{W}$ and the phase shifts matrix $\mathbf{V}$ of the IRS. However, it has  a high computational complexity (i.e. $\mathcal{O}(M^8+N^8)$, according to (\ref{SDR_complex})). To reduce the computational complexity,  a low-complexity SCA-based AO Algorithm is proposed in what follows.

Firstly, by fixing $\mathbf{v}$, problem $\mathrm{(P2)}$ is reduced to
\begin{subequations}\label{P4.1}
\begin{align}
\mathrm{(P4.1):}&\max_{\mathbf{w}}~~~ |\mathbf{v}^H\mathbf{H}_r\mathbf{w}|^2\label{4-1obj}\\
&\text{s. t.}|\mathbf{v}^H\mathbf{H}_b\mathbf{w}|^2+\sigma^2\geq 2^{r_0}(|\mathbf{v}^H\mathbf{H}_e\mathbf{w}|^2+\sigma^2)\label{Rs_4}\\
&~~~~\|\mathbf{w}\|^2 \leq P_s.
\end{align}
\end{subequations}

Note that problem $\mathrm{(P4.1)}$ is still non-convex but objective function (\ref{4-1obj}) is convex. This motivates us to apply SCA method. Since the global lower-bound of a convex function can be expressed by the first-order Taylor expansion at any feasible point \cite{WU_SWIPT_IRS}, the first-order Taylor expansion of $\mathbf{x}^H\mathbf{A}\mathbf{x}$ at point $\tilde{\mathbf{x}}$ is
$\mathbf{x}^H\mathbf{A}\mathbf{x}\geq
2\Re\{\mathbf{x}^H\mathbf{A}\tilde{\mathbf{x}}\}-\tilde{\mathbf{x}}^H\mathbf{A}\tilde{\mathbf{x}}.$
Therefore, problem $\mathrm{(P4.1)}$ can be further translated as
\begin{subequations}\label{P4.1'}
\begin{align}
\mathrm{(P4.1'):}&\max_{\mathbf{w}}~~~ 2\Re\{\mathbf{w}^H\mathbf{H}_{rv}\tilde{\mathbf{w}}\}-\tilde{\mathbf{w}}^H\mathbf{H}_{rv}\tilde{\mathbf{w}}\\
&~~\text{s. t.}~~ 2^{r_0}(\mathbf{w}^H\mathbf{H}_{ev}\mathbf{w}+\sigma^2)\nonumber\\
&~~~~~~~~\leq 2\Re\{\mathbf{w}^H\mathbf{H}_{bv}\tilde{\mathbf{w}}\}-\tilde{\mathbf{w}}^H\mathbf{H}_{bv}\tilde{\mathbf{w}}+ \sigma^2\\
&~~~~~~~~ \mathbf{w}^H\mathbf{w}\leq P_s,
\end{align}
\end{subequations}
where $\mathbf{H}_{iv} = \mathbf{H}_{i}^H\mathbf{v}\mathbf{v}^H\mathbf{H}_i$, $i$ takes $r,e,b$ respectively. $\tilde{\mathbf{w}}$ is the transmit beamforming vector of previous iteration. Subproblem $\mathrm{(P4.1')}$ can be optimally solved by using existing software (e.g. CVX). 

And then by fixing $\mathbf{w}$, problem $\mathrm{(P2)}$ is simplified as
\begin{subequations}\label{P4.2}
\begin{align}\label{4-2obj}
&\mathrm{(P4.2):}\max_{\mathbf{u}}~~~ |\mathbf{u}^H\mathbf{a}+\alpha|^2\\
&~~~~~~\text{s. t.}~~|\mathbf{u}^H\mathbf{b}+\beta|^2+\sigma^2\geq 2^{r_0}(|\mathbf{u}^H\mathbf{c}+\mathbf{\gamma}|^2+\sigma^2) \label{st ube} \\\label{st site}
&~~~~~~~~~~~~|\mathbf{u}_n|=1, \forall n=1\cdots N,
\end{align}
\end{subequations}
where $\mathbf{a}=\mathrm{diag}\{\mathbf{h}_{ih}^H\}\mathbf{G}{\mathbf{w}}$, ${\alpha}=\mathbf{h}_{ah}^H{\mathbf{w}}$, $\mathbf{b}=\mathrm{diag}\{\mathbf{h}_{ib}^H\}\mathbf{G}{\mathbf{w}}$, ${\beta}=\mathbf{h}_{ab}^H{\mathbf{w}}$, $\mathbf{c}=\mathrm{diag}\{\mathbf{h}_{ie}^H\}\mathbf{G}{\mathbf{w}}$, ${\gamma}=\mathbf{h}_{ae}^H{\mathbf{w}}$.
By applying first-order Taylor expansion, the objective function (\ref{4-2obj}) can be expressed as
\begin{align}\label{}
&|\mathbf{u}^H\mathbf{a}+\alpha|^2
\geq2\Re\{\mathbf{u}^H\mathbf{d}\}+c_1,
\end{align}
where $\mathbf{d} = \mathbf{a}\mathbf{a}^H\tilde{\mathbf{u}}+\mathbf{a}\alpha^*$ , $c_1=\alpha\alpha^*-\tilde{\mathbf{u}}^H\mathbf{a}\mathbf{a}^H\tilde{\mathbf{u}}$ and  $\tilde{\mathbf{u}}$ is the phase shifts vector of previous iteration. Similarly, (\ref{st ube}) can be expressed as
\begin{align}\label{A_ini}
\mathbf{u}^H\mathbf{A}\mathbf{u}+2\Re\{\mathbf{u}^H(2^{r_0}\mathbf{c}\gamma^*-\mathbf{b}\beta^*)\}\nonumber\\
\leq \beta\beta^*+\sigma^2-2^{r_0}(\gamma\gamma^*+\sigma^2).
\end{align}
Furthermore $\mathbf{u}^H\mathbf{A}\mathbf{u}$ can be rewritten as\cite{SHEN_Secrecy_IRS}
\begin{equation}\label{AM_S}
\mathbf{u}^H\mathbf{A}\mathbf{u}
\!\leq\! \mathbf{u}^H\mathbf{M}\mathbf{u}\!+\!2\Re\{\mathbf{u}^H(\mathbf{A} \!-\!\mathbf{M})\tilde{\mathbf{u}}\}\!+\!\tilde{\mathbf{u}}^H(\mathbf{M}\!-\!\mathbf{A})\tilde{\mathbf{u}},
\end{equation}\
where $\mathbf{A}=2^{r_0}\mathbf{c}\mathbf{c}^H-\mathbf{b}\mathbf{b}^H$, $\mathbf{M}\succeq \mathbf{A}$. In this paper, we set $\mathbf{M}=\lambda_{max}(\mathbf{A})\mathbf{I}_N$ and $\mathbf{u}^H\mathbf{u} = N$, thus  $\mathbf{u}^H\mathbf{M}\mathbf{u}=N\lambda_{max}(\mathbf{A})$.
 $\lambda_{max}(\mathbf{A})$ denotes the largest eigenvalue of matrix $\mathbf{A}$. Substituting (\ref{AM_S}) into  (\ref{A_ini}), (\ref{st ube}) can be reformulated into
\begin{align}\label{AM_end}
2\Re\{\mathbf{u}^H[(\mathbf{M} -\mathbf{A})\tilde{\mathbf{u}}+(\mathbf{b}\beta^*-2^{r_0}\mathbf{c}\gamma^*)]\}
\geq c_2,
\end{align}
where $c_2=N\lambda_{max}(\mathbf{A})+\tilde{\mathbf{u}}^H(\mathbf{M}-\mathbf{A})\tilde{\mathbf{u}}
+2^{r_0}(\gamma\gamma^*+\sigma^2)-\beta\beta^*-\sigma^2$.

However, problem $\mathrm{(P4.3)}$ is still non-convex due to the constraint (\ref{st site}). It is worth noting that there always exists a non-negative $\mu$ such that $\mathrm{(P4.2)}$ can be formulated into the following equivalent problem.
\begin{subequations}
\begin{align}\label{P4.2'}
\mathrm{(P4.2'):}&\max_{\mathbf{u}} ~~2\Re\{\mathbf{u}^H\mathbf{d}\}
+2\mu\Re\{\mathbf{u}^H\mathbf{f}\}\\
&~~\text{s. t.}~~ (\ref{st site}),
\end{align}
\end{subequations}
where $\mathbf{f}=(\mathbf{M} -\mathbf{A})\tilde{\mathbf{u}}+(\mathbf{b}\beta^*-2^{r_0}\mathbf{c}\gamma^*)$. When the phase shifts $\mathbf{u}$ are equal to $\mathbf{d}+\mu\mathbf{f}$, the objective function is maximized. Therefore, the optimal solution to problem $\mathrm{(P4.2')}$ is
\begin{align}\label{u_solution}
\mathbf{u(\mu)}=e^{j\arg(\mathbf{d}+\mu\mathbf{f})},
\end{align}
where $\mu$ is unknown. Substituting $\mathbf{u(\mu)}$ into constraint (\ref{AM_end}), $2\Re(\mathbf{u(\mu)}\mathbf{f})$ is a monotone increasing function and $\mu$ can be obtained by using bisection search according to the complementary slackness condition $\mu(2\Re(\mathbf{u(\mu)}\mathbf{f})-c_2)=0$. The computing details and proof are similar to that in \cite{PAN_IRS_MIMO} and thus are omitted here for brevity. After bisection search, we obtain the result of $\mu$, which is denoted as $\mu^{'}$, therefore, the optimal solution $\mathbf{u(\mu^{'})}$ is obtained. The proposed low-complexity SCA-based AO scheme is summarized in Algorithm 1.
\begin{algorithm}
\begin{enumerate}
  \item \textbf{Initialization:} $P_s$, $\sigma^2$, $\mathbf{\Theta}^0$, convergence accuracy $\epsilon$ and set $t = 0$.
  \item \textbf{repeat}
  \item ~~~~Set t = t + 1.
  \item ~~~With given $\mathbf{\Theta}^{(t-1)}$ and optimize $\mathbf{w}$ according to $\mathrm{(P4.1)}$ by applying SCA technique.
  \item ~~~~Fix $\mathbf{w}^{(t)}$, calculate the $\theta$ using (\ref{u_solution}) where $\mu$ can be obtained by using bisection search.
  \item ~~~~Set $E_r(t)$ according to (\ref{E_h}).
  \item \textbf{until} $(E_r(t)-E_r(t-1))/E_r(t)<\epsilon$.
\end{enumerate}
\caption{Low-Complexity SCA-based AO Algorithm}\label{algorithm 1}
\end{algorithm}

Because the first-order Taylor expansion is applied in the objective function (\ref{4-1obj}) and (\ref{4-2obj}), the objective value of problem $\mathrm{(P2)}$ is non-decreasing after each iteration. Moreover, similar to SDR-based AO algorithm, the objective value of $\mathrm{(P2)}$ has a finite upper-bound, therefore, Algorithm 1 is guaranteed to converge.
\subsection{Complexity Analysis}
In this section, we will calculate  the complexities of the two proposed methods and make a comparison with existing  methods. The total complexity of the proposed SDR-based AO algorithm  without Gaussian randomization is\cite{complexity}
\begin{align}
&\mathcal{O}\{D[\sqrt{(2\!+\!M)}\big(M^2(2\!+\!M^3)\!+\!M^4(2\!+\!M^2)\!+\!M^8\big)+\nonumber\\
&\sqrt{(2N\!+\!2)}\big(N^2(N^3\!+\!N\!+\!2)+N^4(N^2\!+\!N\!+\!2)+N^8\big)]\},\label{SDR_complex}
\end{align}
where $D$ denotes the number of alternating iterations.
%

The complexity of the proposed SCA-based AO algorithm is given by
\begin{align}
\mathcal{O}\{L[L_1\big(2M(5^2+(M+1)^2)+M^3\big)~~~~~~~~~~~~\nonumber\\
+\big(L_2N^3\log_2((\lambda_{max}-\lambda_{min})/{\varepsilon})\big)]\},
\end{align}
where $L$ denotes the maximum numbers of alternating iterations. $L_1$ and $L_2$ denote the iterative number of SCA in subproblems $\mathrm{(P4.1)}$ and $\mathrm{(P4.2)}$, respectively. $\lambda_{max}$, $\lambda_{min}$ and $\varepsilon$ are the upper-bound, lower-bound, and the accuracy of bisection method, respectively. $\log_2((\lambda_{max}-\lambda_{min})/{\varepsilon})$ is the maximum number of bisection search.
Obviously, although the proposed SCA-based AO algorithm involves two-level iteration, the highest order of computational complexity is only $M^3$ and $N^3$ FLOPS compared to $M^8$ and $N^8$  FLOPS of the proposed SDR-based AO algorithm. Therefore, the computational complexity of the SCA-based AO algorithm is much lower than that of SDR-based AO Algorithm, especially in massive IRS or massive MIMO  scenario.
\section{Simulation and Discussion}
In this section, we evaluate the EHR performance of the proposed method by numerical simulation. Two benchmark schemes are used: 1) Random phase shifts, which means $\mathbf{\theta}_n~(n=1\cdots N)$ is randomly  chosen from  the interval $\mathrm{[0,2\pi)}$. 2) Without IRS, i.e. $\mathbf{\Theta}=0$. In our simulation, it is assumed that all the channels experience Rayleigh fading. The reference path loss is 30dB per 1m. Since the IRS is often placed to avoid  blocking the  signal from the AP, thus, the path loss exponent of AP-IRS, IRS-Bob/EHR/EVE is set to 2, whereas the path loss exponent of AP-Bob/EHR/EVE is set  to 3. In addition, the distances of AP-EHR, AP-EVE, AP-Bob are 6m, 85m, 220m, respectively. And AP-IRS channel is assumed to be line-of-sight (LoS) link and the associated  distance is 8m. Other simulation parameters are set as follows: $\sigma^2=-70\mathrm{dBm}, \zeta=0.5$, and  $\mathrm{P}_s=15\mathrm{W}$.
\begin{figure}[htb]
  \centering
  \includegraphics[width=0.45\textwidth]{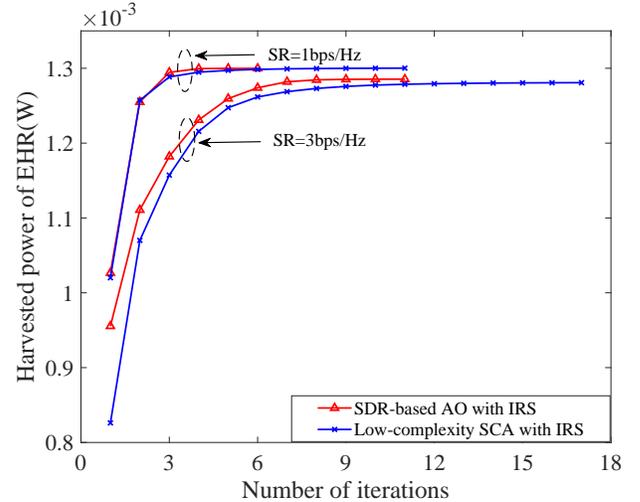}
  \caption{Harvested power of EHR versus the number of iterations.}
  \label{Fig2_Convergence}
\end{figure}

Fig.~\ref{Fig2_Convergence} demonstrates the convergence of the proposed methods for SR being 1bps/Hz and 3bps/Hz, respectively. It is seen from Fig.~\ref{Fig2_Convergence} that the two proposed AO methods could converge rapidly to the power ceils within about ten iterations. This verifies the feasibility of the algorithm. After convergence, the two proposed schemes may achieve an excellent harvested power improvement over initial phases.

\begin{figure}[htb]
  \centering
  \includegraphics[width=0.45\textwidth]{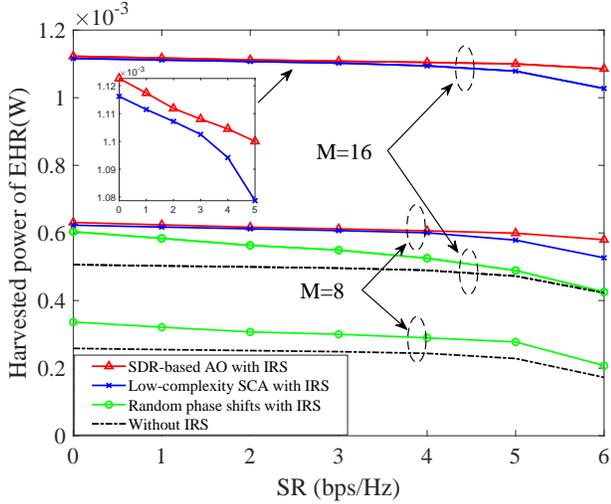}
  \caption{Harvested power of EHR versus SR.}
  \label{Fig3_difM}
\end{figure}

Fig.~\ref{Fig3_difM} illustrates the curves of harvested power of EHR versus SR for $N$ = 50.
From this figure, it is observed that the harvested power of the proposed schemes decreases with increase in SR. Compared with conventional scheme, the two proposed methods with the aid of IRS achieve an approximate doubled harvested power. This is because the IRS provides a new degree of freedom and diversity gain to enhance the harvested power of EHR by optimizing the phase shifts at IRS. Moreover, the proposed two methods perform  much better than conventional method without IRS and random-phase-shifts method with IRS. When SR$<$4bps/Hz, the proposed SCA method approaches the proposed SDR one in terms of  harvested power. When SR$>$4bps/Hz, the former is slightly worse than the latter. Additionally, increasing the number of antennas at AP accordingly improves the harvested power at EHR.

\begin{figure}[htb]
  \centering
  \includegraphics[width=0.45\textwidth]{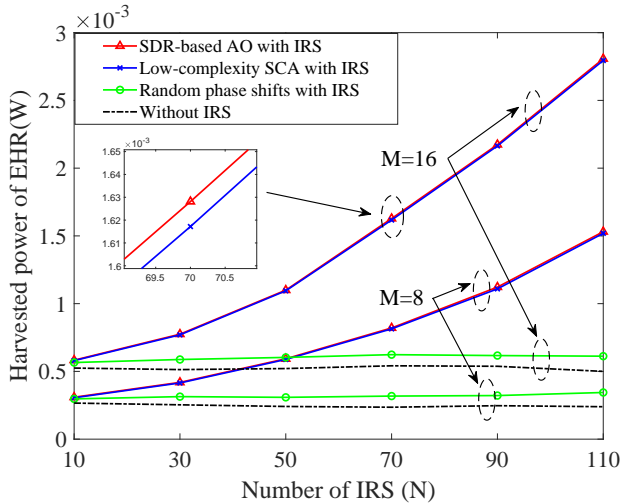}
  \caption{Harvested power of EHR versus N.}
  \label{Fig4_difN}
\end{figure}

Fig.~\ref{Fig4_difN} plots the harvested power of EHR versus $N$ with SR being 1bps/Hz. Obviously, the proposed methods are still better than existing methods. As $N$ increases, the harvested power at EHR increases gradually. The main reason  is that with more reflecting elements, the more new degrees of freedom achieved by IRS. Consistent with Fig.~\ref{Fig3_difM}, the random-phase-shifts method with IRS has a minor improvement over that without IRS.
\section{Conclusion}
In this paper, we have presented an investigation of secure transmit beamforming and phase shifting at IRS in a secure IRS-assisted MISO-SWIPT network to maximize the harvested power at EHR. Two alternating iterative algorithms SDR and SCA were proposed to address the non-convex optimization problem. Using a much lower-complexity, the proposed SCA method can achieve the same performance as the proposed SDR method. More importantly, with the help of IRS, the proposed two methods double the harvested power compared to existing methods. 

\ifCLASSOPTIONcaptionsoff
  \newpage
\fi
\bibliographystyle{IEEEtran}
\bibliography{IEEEfull,cite}

\begin{thebibliography}{10}
\providecommand{\url}[1]{#1}
\csname url@samestyle\endcsname
\providecommand{\newblock}{\relax}
\providecommand{\bibinfo}[2]{#2}
\providecommand{\BIBentrySTDinterwordspacing}{\spaceskip=0pt\relax}
\providecommand{\BIBentryALTinterwordstretchfactor}{4}
\providecommand{\BIBentryALTinterwordspacing}{\spaceskip=\fontdimen2\font plus
\BIBentryALTinterwordstretchfactor\fontdimen3\font minus
  \fontdimen4\font\relax}
\providecommand{\BIBforeignlanguage}[2]{{%
\expandafter\ifx\csname l@#1\endcsname\relax
\typeout{** WARNING: IEEEtran.bst: No hyphenation pattern has been}%
\typeout{** loaded for the language `#1'. Using the pattern for}%
\typeout{** the default language instead.}%
\else
\language=\csname l@#1\endcsname
\fi
#2}}
\providecommand{\BIBdecl}{\relax}
\BIBdecl

\bibitem{wu_TWC}
Q.~{Wu} and R.~{Zhang}, ``Intelligent reflecting surface enhanced wireless
  network via joint active and passive beamforming,'' \emph{IEEE Trans.
  Wireless Commun.}, pp. 1--1, 2019.

\bibitem{Basar_IRS}
E.~{Basar}, M.~{Di Renzo}, J.~{De Rosny}, M.~{Debbah}, M.~{Alouini}, and
  R.~{Zhang}, ``Wireless communications through reconfigurable intelligent
  surfaces,'' \emph{IEEE Access}, vol.~7, pp. 116\,753--116\,773, 2019.

\bibitem{wu_one}
Q.~{Wu} and R.~{Zhang}, ``Intelligent reflecting surface enhanced wireless
  network: Joint active and passive beamforming design,'' in \emph{in Proc.
  IEEE Global Commun.Conf. (GLOBECOM)}, Dec 2018, pp. 1--6.

\bibitem{Energy_Efficiency}
C.~{Huang}, A.~{Zappone}, G.~C. {Alexandropoulos}, M.~{Debbah}, and C.~{Yuen},
  ``Reconfigurable intelligent surfaces for energy efficiency in wireless
  communication,'' \emph{IEEE Trans. Wireless Commun.}, vol.~18, no.~8, pp.
  4157--4170, Aug 2019.

\bibitem{GUO_weight_sum_rate}
H.~{Guo}, Y.-C. {Liang}, J.~{Chen}, and E.~G. {Larsson}, ``Weighted sum-rate
  optimization for intelligent reflecting surface enhanced wireless networks,''
  [online] Available: http://arxiv.org/abs/1905.07920.

\bibitem{chenXM_secury}
X.~{Chen}, D.~W.~K. {Ng}, W.~H. {Gerstacker}, and H.~{Chen}, ``A survey on
  multiple-antenna techniques for physical layer security,'' \emph{IEEE
  Communications Surveys Tutorials}, vol.~19, no.~2, pp. 1027--1053,
  Secondquarter 2017.

\bibitem{wangMH_secury}
H.~{Wang}, Q.~{Yin}, and X.~{Xia}, ``Distributed beamforming for physical-layer
  security of two-way relay networks,'' \emph{IEEE Trans. Signal Process},
  vol.~60, no.~7, pp. 3532--3545, July 2012.

\bibitem{zhangN_secury}
N.~{Zhao}, F.~R. {Yu}, M.~{Li}, and V.~C.~M. {Leung}, ``Anti-eavesdropping
  schemes for interference alignment {(IA)}-based wireless networks,''
  \emph{IEEE Trans. Wireless Commun.}, vol.~15, no.~8, pp. 5719--5732, Aug
  2016.

\bibitem{zhou_UAV}
X.~{Zhou}, Q.~{Wu}, S.~{Yan}, F.~{Shu}, and J.~{Li}, ``{UAV}-enabled secure
  communications: Joint trajectory and transmit power optimization,''
  \emph{IEEE Trans. Veh.Technol.}, vol.~68, no.~4, pp. 4069--4073, April 2019.

\bibitem{cui_security}
M.~{Cui}, G.~{Zhang}, and R.~{Zhang}, ``Secure wireless communication via
  intelligent reflecting surface,'' \emph{IEEE Wireless Communications
  Letters}, vol.~8, no.~5, pp. 1410--1414, Oct 2019.

\bibitem{SHEN_Secrecy_IRS}
H.~{Shen}, W.~{Xu}, W.~{Xu}, S.~{Gong}, Z.~{He}, and C.~{Zhao}, ``Secrecy rate
  maximization for intelligent reflecting surface assisted multi-antenna
  communications,'' \emph{IEEE Commun. Lett.}, vol.~23, no.~9, pp. 1488--1492,
  Sep. 2019.

\bibitem{XU_SWIPT_SEC}
J.~{Xu}, L.~{Liu}, and R.~{Zhang}, ``Multiuser {MISO} beamforming for
  simultaneous wireless information and power transfer,'' \emph{IEEE Trans.
  Signal Process.}, vol.~62, no.~18, pp. 4798--4810, Sep. 2014.

\bibitem{WU_SWIPT_IRS}
Q.~{Wu} and R.~{Zhang}, ``Weighted sum power maximization for intelligent
  reflecting surface aided {SWIPT},'' [online] Available:
  http://arxiv.org/abs/1907.05558.

\bibitem{PAN_IRS_MIMO}
C.~{Pan}, H.~{Ren}, K.~{Wang}, M.~{Elkashlan}, A.~{Nallanathan}, J.~{Wang}, and
  L.~{Hanzo}, ``Intelligent reflecting surface aided {MIMO} broadcasting for
  simultaneous wireless information and power transfer,'' [online] Available:
  http://arxiv.org/abs/1908.04863.

\bibitem{wuXM_Secury}
F.~{Shu}, X.~{Wu}, J.~{Hu}, J.~{Li}, R.~{Chen}, and J.~{Wang}, ``Secure and
  precise wireless transmission for random-subcarrier-selection-based
  directional modulation transmit antenna array,'' \emph{IEEE J. Sel. Areas
  Commun.}, vol.~36, no.~4, pp. 890--904, April 2018.

\bibitem{complexity}
K.~{Wang}, A.~M. {So}, T.~{Chang}, W.~{Ma}, and C.~{Chi}, ``Outage constrained
  robust transmit optimization for multiuser {MISO} downlinks: Tractable
  approximations by conic optimization,'' \emph{IEEE Trans. Signal Process.},
  vol.~62, no.~21, pp. 5690--5705, Nov 2014.

\end{thebibliography}
\end{document}